\documentclass[10pt,conference]{IEEEtran}
\IEEEoverridecommandlockouts
\usepackage{cite}
\usepackage{amsmath,amssymb,amsfonts}
\usepackage{algorithmic}
\usepackage{graphicx}
\usepackage{textcomp}
\usepackage{xcolor}
\def\BibTeX{{\rm B\kern-.05em{\sc i\kern-.025em b}\kern-.08em
    T\kern-.1667em\lower.7ex\hbox{E}\kern-.125emX}}
    
\usepackage[hang, flushmargin]{footmisc}
\usepackage[colorlinks, allcolors=blue]{hyperref}
\usepackage[stretch=10]{microtype}
\usepackage{enumitem}
\usepackage{varwidth}
\usepackage{tasks}
\usepackage{soul}
\usepackage{blindtext}
\usepackage{setspace}
\usepackage{array}
\usepackage{ragged2e}
\usepackage{multirow}
\usepackage{booktabs}
\usepackage{float}
\usepackage{caption}
\usepackage{subcaption}
\usepackage[most]{tcolorbox}
\usepackage{placeins}
\usepackage{arydshln}
\usepackage{balance}
\usepackage{pgfplots}

\newtcolorbox[auto counter]{prompt}[1][]{
  enhanced,
  boxrule=0.2pt,
  boxsep=-0.2mm,
  colback=blue!5!white!50,
  colframe=blue!30!black,
  breakable,
  #1
}

\definecolor{bg}{HTML}{F8F9FB}  
\definecolor{bgc}{HTML}{FCF6E4}
\definecolor{rowcolor}{HTML}{ECEFF4}  
\definecolor{highlightY}{HTML}{FAE6A2}

\newcommand{\logtext}[1]{``\textls[-50]{\texttt{#1}}"}

\newcommand{\subscript}[2]{#1#2.}

\def\BibTeX{{\rm B\kern-.05em{\sc i\kern-.025em b}\kern-.08em
    T\kern-.1667em\lower.7ex\hbox{E}\kern-.125emX}}
\begin{document}

\title{Log Parsing: How Far Can ChatGPT Go?}

\author{\IEEEauthorblockN{
Van-Hoang Le$^{1}$ and
Hongyu Zhang$^{2}$\IEEEauthorrefmark{2}\thanks{\IEEEauthorrefmark{2}Hongyu Zhang is the corresponding author.}
}
\IEEEauthorblockA{$^1$School of Information and Physical Sciences, The University of Newcastle, Australia}
\IEEEauthorblockA{$^2$School of Big Data and Software Engineering, Chongqing University, China}
\IEEEauthorblockA{vanhoang.le@uon.edu.au, hyzhang@cqu.edu.cn}
}

\maketitle

\begin{abstract}

Software logs play an essential role in ensuring the reliability and maintainability of large-scale software systems, as they are often the sole source of runtime information. Log parsing, which converts raw log messages into structured data, is an important initial step towards downstream log analytics. In recent studies, ChatGPT, the current cutting-edge 
large language model (LLM), has been widely applied to a wide range of software engineering tasks. However, its performance in automated log parsing remains unclear. In this paper, we evaluate ChatGPT’s ability to undertake log parsing by addressing two research questions. (1) Can ChatGPT effectively parse logs? (2) How does ChatGPT perform with different prompting methods? 
Our results show that ChatGPT can achieve promising results
for log parsing with appropriate prompts, especially with
few-shot prompting.
Based on our findings, we outline several challenges and opportunities for ChatGPT-based log parsing.
\end{abstract}

\section{Introduction}
Large-scale software-intensive systems, such as cloud computing and big data systems, generate a large 
amount of logs for troubleshooting purposes.
Log messages are produced during software runtime by the logging statements in source code. 
They record system events and dynamic runtime information,  
which can help developers and operators understand system behavior and perform system diagnostic tasks, such as anomaly detection~\cite{le2022log, du2017deeplog, zhang2019robust, le2021log}, failure prediction~\cite{das2018desh, russo2015mining}, and failure diagnosis~\cite{lu2017log, gurumdimma2016crude}.


Log parsing is an important initial step of many downstream log-based system diagnostic tasks. Through log parsing, free-text raw log messages are converted into a stream of structured events~\cite{du2016spell, he2017drain, zhu2019tools, guideline2022}.
To achieve better log parsing accuracy, many data-driven approaches, such as those based on clustering~\cite{tang2011logsig, shima2016length}, frequent pattern mining~\cite{dai2020logram, nagappan2010abstracting}, and heuristics~\cite{he2017drain, du2016spell, li2020_swisslog}, have been proposed to automatically distinguish the constant and variable parts of log messages~\cite{guideline2022, zhu2019tools, le2023log}. 
Recent studies adopt pre-trained language models for representing~\cite{le2021log, le2023log, tao2021logstamp} log data. 
However, these methods still require either training models from scratch~\cite{liu2022uniparser} or tuning a pre-trained language model with labelled data~\cite{le2021log,le2023log}, which could be impractical due to the scarcity of computing resources and labelled data.

More recently, 
large language models (LLMs) such as ChatGPT~\cite{chatgpt2023} has been applied to a variety of software engineering tasks 
and achieved satisfactory performance 
~\cite{gao2023makes, cao2023study}. 
However, it is unclear whether or not ChatGPT can effectively perform automated log parsing. More research is needed to determine its capabilities in this important area.
Therefore, in this paper, we conduct a preliminary evaluation of ChatGPT for log parsing. 



More specifically, we design appropriate prompts to guide ChatGPT to understand the log parsing task and extract the log event/template from the input log messages.
We then compare the performance of ChatGPT with that of SOTA (state-of-the-art) log parsers in zero-shot scenario. We also examine the performance of ChatGPT with a few log parsing demonstrations (few-shot scenarios). Finally, we analyze the performance of ChatGPT 
to explore its potential in log parsing. Our experimental results show that ChatGPT can achieve promising results for log parsing with appropriate prompts, especially with few-shot prompting. 
We also outline several challenges and opportunities for ChatGPT-based log parsing.

In summary, the major contributions of this work are as follows:
\begin{itemize}[leftmargin=12pt]
 
    \item To the best of our knowledge, we are the first to investigate and analyze ChatGPT's ability to undertake log parsing. 
    \item 
We evaluate ChatGPT-based log parsing on widely-used log datasets and compare it with SOTA log parsers.
   \item Based on the findings, we outline several challenges and prospects for ChatGPT-based log parsing.
    
\end{itemize}

\section{Background}
\subsection{Log Data}
\begin{figure}[h]
    \centering
    \includegraphics[width=\linewidth]{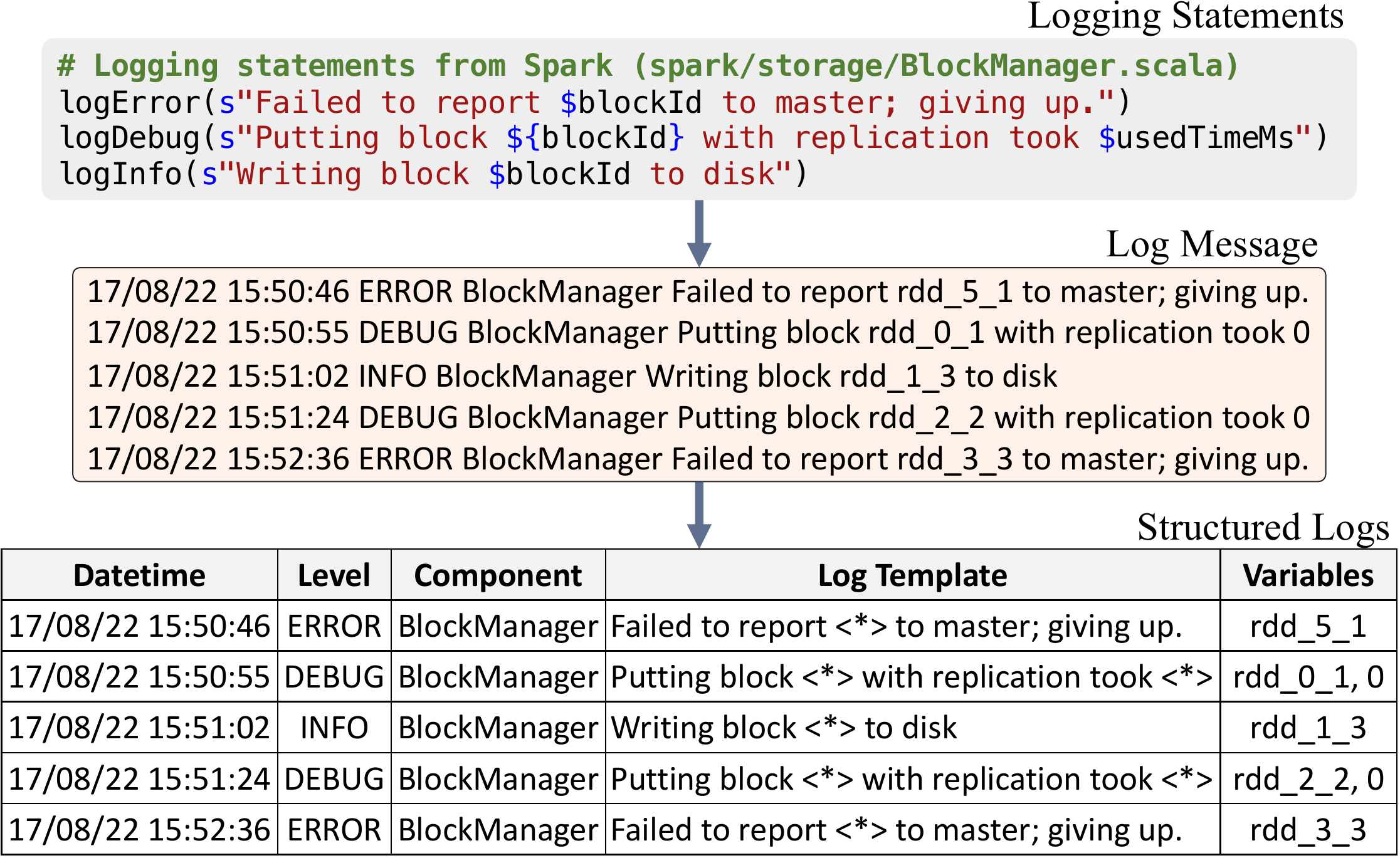}
    \caption{An example of log data and log parsing from Spark}
    \label{fig:log_example}
\end{figure}

Large and complex software-intensive systems often produce a large amount of log data for troubleshooting purposes during system operation. Log data records the system’s events and internal states during runtime. 
Figure~\ref{fig:log_example} shows a snippet of log data generated by Spark.

A log message usually contains a header that is automatically produced by the logging framework, including information such as component and verbosity level. The log message body (log message for short) typically consists of two parts: \textit{1) Template} - constant strings (or keywords) describing the system events; \textit{2) Parameters/Variables} - dynamic variables, which 
reflect specific runtime status. Figure~\ref{fig:log_example} shows an example of logs produced by Spark, where the header (including Datetime, Component, and Level) is generated by the logging framework and is generally easy to extract. The log event/template \logtext{Putting block <*> with replication took <*>} associated with parameters (e.g., \logtext{rdd\_0\_1}, \logtext{0}), in contrast, is often difficult to identify.

\subsection{Log Parsing}
Log parsing typically serves as the first step toward automated log analytics.
It aims at parsing each log message into a specific log event/template and extracting the corresponding parameters. 
To achieve better performance compared to traditional regular expression-based log parsers, many data-driven approaches have been proposed to automatically distinguish template and parameter parts. These data-driven approaches can be categorized into several groups. 1) \textit{Similarity-based clustering:} LKE~\cite{fu2009execution}, LogSig~\cite{tang2011logsig}, and LenMa~\cite{shima2016length} compute distances between two log messages or their signature to cluster them based on similarity. 2) \textit{Frequent pattern mining:} SLCT~\cite{vaarandi2003data}, LFA~\cite{nagappan2010abstracting}, and Logram~\cite{dai2020logram} leverage frequent patterns of token position or $n$-gram information to extract log templates that appear constantly across log messages. \textit{3) Heuristics-based searching:} Drain~\cite{he2017drain}, Spell~\cite{du2016spell}, SwissLog~\cite{li2020_swisslog}, and SPINE~\cite{wang2022spine} utilize a tree structure to parse logs into multiple templates. \textit{4) Deep learning based parsing:} UniParser~\cite{liu2022uniparser} formulates log parsing as a token classification problem and LogPPT~\cite{le2023log} leverages language models to perform log parsing in few-shot scenarios.

Structured log data, obtained after log parsing, can be used for many downstream log analytics tasks, such as
anomaly detection~\cite{he2016experience, du2017deeplog, zhang2019robust}, failure prediction~\cite{das2018desh, li2021prefix}, and failure diagnosis~\cite{gurumdimma2016crude, lu2017log}.


\section{Study Design}
\subsection{Research Questions}
We aim at answering the following research questions through experimental evaluation:
\begin{enumerate}[label=\textbf{\subscript{RQ}{{\arabic*}}}, leftmargin=1cm]
    \item Can ChatGPT effectively perform log parsing?
    \item How does ChatGPT perform with different prompting methods? 
\end{enumerate}


RQ1 is to evaluate the effectiveness of ChatGPT in log parsing. To answer RQ1, we provide a basic definition of log parsing (i.e., abstracting the dynamic variables in logs~\cite{li2023did}) and ask ChatGPT to extract the log template for one log message per request by using the following prompt template (where the slot \textbf{`[LOG]'} indicates the input log message):

\begin{prompt}[label=pt:zero-shot-1, title=\strut \textbf{Prompt Template (PT\thetcbcounter)}]
\small
You will be provided with a log message delimited by backticks. You must abstract variables with `\{placeholders\}' to extract the corresponding template. Print the input log's template delimited by backticks.\\
\\Log message: `\sethlcolor{highlightY}\hl{[LOG]}'
\end{prompt}

RQ2 aims at investigating the impact of prompting methods 
on ChatGPT-based log parsing. 
Specifically, we evaluate the performance of ChatGPT under two experimental settings:
\begin{enumerate}[wide, labelwidth=!, labelindent=0pt]
    \item \textbf{Few-shot scenarios:} Since log data is heterogeneous, we follow a recent study~\cite{le2023log} to provide a few demonstrations (1, 2, and 4) of log parsing when applying ChatGPT to log data. Specifically, we use the following prompt template to ask ChatGPT to extract the log template of an input log:
\begin{prompt}[label=pt:fewshot, title=\strut \textbf{Prompt Template (PT\thetcbcounter)}]
\small
You will be provided with a log message delimited by backticks. You must abstract variables with `\{placeholders\}' to extract the corresponding template.

For example:\\ 
The template of `[DEMO\_LOG1]' is `[TEMPLATE1]'.\\
The template of `[DEMO\_LOG2]' is `[TEMPLATE2]'.\\
...
\\
Print the input log's template delimited by backticks.\\
\\Log message: `\sethlcolor{highlightY}\hl{[LOG]}'
\end{prompt}
    \item \textbf{Different prompts:} We evaluate the impact of different prompts on log parsing with ChatGPT. Specifically, along with PT\ref{pt:zero-shot-1}, we further evaluate a \textit{simple} (PT\ref{pt:zero-shot-simple}) and an \textit{enhanced} (PT\ref{pt:zero-shot-enhance}) prompt as follows:
    
\begin{prompt}[label=pt:zero-shot-simple, title=\strut \textbf{Prompt Template - Simple (PT\thetcbcounter)}]
\small
You will be provided with a log message delimited by backticks. Please extract the log template from this log message:\\
`\sethlcolor{highlightY}\hl{[LOG]}'
\end{prompt}

\begin{prompt}[label=pt:zero-shot-enhance, title=\strut \textbf{Prompt Template - Enhance (PT\thetcbcounter)}]
\small
You will be provided with a log message delimited by backticks. You must identify and abstract all the dynamic variables in logs with `\{placeholders\}` and output a static log template.
Print the input log's template delimited by backticks.\\
\\Log message: `\sethlcolor{highlightY}\hl{[LOG]}'
\end{prompt}
\end{enumerate}


\subsection{Benchmark and Setting}
We conduct experiments on 16 datasets originated from the LogPai benchmark~\cite{he2020loghub, zhu2019tools}. This benchmark covers log data from various systems, including distributed systems, supercomputers, operating systems, mobile systems, server applications, and standalone software. Since there are multiple errors in the original benchmark, in this paper, we use the corrected version of this benchmark~\cite{khan2022guidelines} in our evaluation. Each dataset contains 2,000 manually labelled log messages.

We build a pipeline for our experiments using the ChatGPT API based on the gpt-3.5-turbo model released by OpenAI~\cite{chatgpt2023}. To avoid bias from model updates, we use a snapshot of gpt-3.5-turbo from March 2023~\cite{gpt-3.5-turbo}.

\begin{table*}[htbp]
\caption{Comparison with existing log parsers in zero-shot scenario}
\label{tab:rq_acc}
\centering
\footnotesize
\setlength{\tabcolsep}{2.5pt}
\renewcommand{\arraystretch}{1.3}
\begin{tabular}{c|ccc|ccc|ccc|ccc|ccc|ccc}
\hline
\multicolumn{1}{l|}{\multirow{2}{*}{}} & \multicolumn{3}{c|}{\textbf{AEL}} & \multicolumn{3}{c|}{\textbf{Spell}} & \multicolumn{3}{c|}{\textbf{Drain}} & \multicolumn{3}{c|}{\textbf{Logram}} & \multicolumn{3}{c|}{\textbf{SPINE}} & \multicolumn{3}{c}{\textbf{ChatGPT}} \\ \cline{2-19} 
\multicolumn{1}{l|}{} & GA & MLA & ED & GA & MLA & ED & \multicolumn{1}{c}{GA} & \multicolumn{1}{c}{MLA} & \multicolumn{1}{c|}{ED} & \multicolumn{1}{c}{GA} & \multicolumn{1}{c}{MLA} & \multicolumn{1}{c|}{ED} & \multicolumn{1}{c}{GA} & \multicolumn{1}{c}{MLA} & \multicolumn{1}{c|}{ED} & \multicolumn{1}{c}{GA} & \multicolumn{1}{c}{MLA} & \multicolumn{1}{c}{ED} \\ \hline
HDFS & 0.998 & 0.625 & 0.943 & \textbf{1} & 0.301 & 1.386 & 0.998 & 0.626 & 0.940 & 0.930 & 0.005 & 19.297 & 0.866 & 0.499 & 7.980 & 0.960 & \textbf{0.939} & \textbf{0.062} \\
Hadoop & 0.869 & 0.262 & 14.576 & 0.778 & 0.113 & 23.967 & 0.948 & 0.269 & 15.399 & 0.451 & 0.113 & 26.531 & \textbf{0.950} & 0.279 & 16.447 & 0.795 & \textbf{0.525} & \textbf{11.017} \\
Spark & 0.905 & 0.360 & 3.197 & 0.905 & 0.321 & 5.465 & 0.920 & 0.360 & 2.629 & 0.282 & 0.259 & 7.532 & \textbf{0.925} & 0.337 & 4.816 & \textbf{0.925} & \textbf{0.922} & \textbf{0.596} \\
Zookeeper & 0.921 & 0.496 & \textbf{2.672} & 0.964 & 0.452 & 3.188 & 0.967 & 0.497 & 2.288 & 0.724 & 0.474 & 5.534 & \textbf{0.989} & \textbf{0.502} & 3.541 & 0.667 & 0.233 & 5.460 \\
BGL & 0.957 & 0.344 & 5.057 & 0.787 & 0.197 & 7.982 & \textbf{0.963} & 0.344 & \textbf{4.973} & 0.587 & 0.125 & 10.021 & 0.923 & 0.376 & 5.081 & 0.878 & \textbf{0.790} & 5.258 \\
HPC & 0.903 & 0.678 & 0.959 & 0.654 & 0.530 & 4.630 & 0.887 & 0.654 & 1.534 & 0.911 & 0.665 & 2.278 & \textbf{0.945} & \textbf{0.667} & \textbf{1.980} & 0.807 & 0.497 & 3.498 \\
Thunderb & 0.941 & 0.036 & 14.731 & 0.844 & 0.027 & 15.684 & \textbf{0.955} & 0.047 & 14.632 & 0.554 & 0.004 & 16.208 & 0.665 & 0.051 & 18.331 & 0.568 & \textbf{0.808} & \textbf{5.933} \\
Windows & 0.690 & 0.153 & 10.767 & 0.989 & 0.004 & 3.200 & \textbf{0.997} & \textbf{0.462} & \textbf{4.966} & 0.694 & 0.141 & 6.700 & 0.684 & 0.151 & 12.379 & 0.686 & 0.301 & 17.623 \\
Linux & 0.405 & 0.174 & 15.633 & 0.152 & 0.088 & 16.256 & 0.422 & 0.177 & 15.534 & 0.186 & 0.124 & 17.857 & 0.545 & 0.108 & 11.145 & \textbf{0.910} & \textbf{0.635} & \textbf{3.328} \\
Android & 0.773 & 0.393 & \textbf{9.396} & 0.863 & 0.150 & 12.574 & 0.831 & 0.548 & 6.940 & 0.742 & 0.278 & 17.734 & \textbf{0.938} & 0.181 & 14.630 & 0.711 & \textbf{0.549} & 10.763 \\
HealthApp & 0.568 & 0.163 & 19.066 & 0.639 & 0.152 & 8.468 & 0.780 & 0.231 & 18.476 & 0.267 & 0.112 & 15.814 & \textbf{0.983} & 0.446 & 5.320 & 0.898 & \textbf{0.628} & \textbf{6.560} \\
Apache & \textbf{1} & 0.694 & 10.218 & \textbf{1} & 0.694 & 10.234 & \textbf{1} & 0.694 & 10.218 & 0.313 & 0.007 & 12.315 & \textbf{1} & 0.276 & 11.036 & \textbf{1} & \textbf{1} & \textbf{0} \\
Proxifier & 0.495 & 0.495 & 10.207 & 0.527 & 0.478 & 12.842 & \textbf{0.527} & \textbf{0.504} & \textbf{10.138} & 0.504 & 0 & 27.222 & 0.049 & 0.016 & 14.198 & 0.001 & 0.014 & 27.025 \\
OpenSSH & 0.537 & 0.246 & \textbf{4.976} & 0.556 & 0.191 & 7.331 & \textbf{0.789} & \textbf{0.508} & 7.543 & 0.611 & 0.298 & 6.220 & 0.676 & 0.253 & 8.018 & 0.659 & 0.170 & 7.854 \\
OpenStack & 0.758 & 0.019 & 19.559 & 0.764 & 0 & 30.400 & \textbf{0.733} & 0.019 & 30.759 & 0.326 & 0 & 64.057 & 0.384 & 0.011 & 48.025 & 0.449 & \textbf{0.433} & \textbf{7.440} \\
Mac & 0.764 & 0.169 & \textbf{18.902} & 0.757 & 0.033 & 23.390 & \textbf{0.787} & 0.230 & 20.365 & 0.568 & 0.182 & 21.517 & 0.761 & 0.204 & 19.334 & 0.619 & 0.248 & 25.530 \\ \hline
Average & 0.780 & 0.331 & \multicolumn{1}{l|}{10.053} & 0.761 & 0.233 & \multicolumn{1}{l|}{11.687} & \textbf{0.844} & 0.385 & 10.458 & 0.540 & 0.174 & 17.302 & 0.767 & 0.272 & 12.641 & 0.721 & \textbf{0.543} & \textbf{8.621} \\ \bottomrule
\multicolumn{19}{l}{%
  \begin{minipage}{.8\linewidth}%
    Note: \textit{Thunderb} denotes Thunderbird; For Edit Distance (ED), the lower is the better.
  \end{minipage}%
}\\
\end{tabular}
\end{table*}

\subsection{Baselines}
We compare our proposed method with five state-of-the-art log parsers, including AEL~\cite{jiang2008abstracting}, Spell~\cite{du2016spell}, Drain~\cite{he2017drain}, Logram~\cite{dai2020logram}, and SPINE~\cite{wang2022spine}.
We choose these five parsers in our evaluation since their source code is publicly available; and a prior study~\cite{zhu2019tools, khan2022guidelines} finds that these parsers have high accuracy and efficiency among the evaluated log parsing methods. For SPINE, we use the source code provided by its authors. For other baselines, we adopt the implementation of these methods from their replication packages~\cite{logparser2022, guideline2022}.

\subsection{Evaluation Metrics}
Following recent studies\cite{guideline2022, liu2022uniparser, nedelkoski2020self-parsing}, we apply three metrics to comprehensively evaluate the effectiveness of log parsing, including:
\begin{itemize}[leftmargin=12pt]
    \item \textbf{Group Accuracy (GA):} Group Accuracy~\cite{zhu2019tools} is the most commonly used metric for log parsing. The GA metric is defined as the ratio of “correctly parsed” log messages over the total number of log messages, where a log message is considered “correctly parsed” if and only if it is grouped with other log messages consistent with the ground truth.
    \item \textbf{Message Level Accuracy (MLA):} The Message Level Accuracy~\cite{liu2022uniparser} metric is defined as the ratio of “correctly parsed” log messages over the total number of log messages, where a log message is considered to be “correctly parsed” if and only if every token of the log message is correctly identified as template or variable.
    \item \textbf{Edit Distance (ED):} Edit Distance~\cite{nedelkoski2020self-parsing} measure the accuracy of log parsers in terms of lexical similarity between parsed results and ground truth, by computing the distance between parsed templates and ground truth templates.
\end{itemize}

\section{Results}
\subsection{The Effectiveness of ChatGPT in Log Parsing}

\begin{figure}[h]
    \centering
    \vspace{-6pt}
    \includegraphics[width=\linewidth]{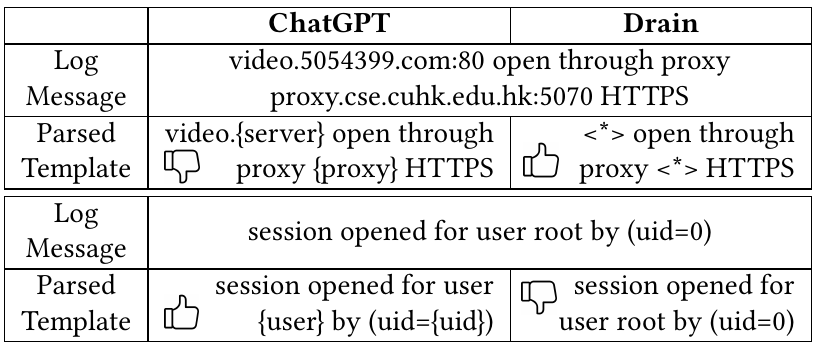}
    \vspace{-6pt}
    \caption{Examples of log parsing with ChatGPT}
    \vspace{-6pt}
    \label{fig:chatgpt-examples}
\end{figure}

In this RQ, we evaluate the performance of ChatGPT-based log parsing in the zero-shot scenario. We compare the results of ChatGPT with baselines in terms of GA, MLA, and ED. Table~\ref{tab:rq_acc} shows the results. For each dataset, the best accuracy is highlighted in boldface.
The results show that, in terms of GA, ChatGPT achieves the best accuracy on three out of 16 datasets. On average, it achieves a GA of 0.721, which outperforms the average result of Logram by 33.5\%, and is 0.85x of that of the best baseline Drain.
Regarding MLA and ED, ChatGPT significantly outperforms the baselines with an improvement of 41.0\% to 212.1\% in MLA and 14.2\% to 50.2\% in ED. Specifically, it achieves the best results on 10 out of 16 datasets in terms of MLA and 8 out of 16 datasets in terms of ED.
The results indicate that ChatGPT is able to distinguish variable and content tokens in log messages, as reflected by the high MLA values.
However, there is much log-specific information (such as domain URLs, API endpoint addresses, block/task IDs, etc), which varies a lot across log data. 
ChatGPT 
has difficulties in correctly recognizing these log-specific information, 
leading to lower GA values. 
Figure~\ref{fig:chatgpt-examples} shows some examples of log templates generated by ChatGPT and Drain. We can see that ChatGPT correctly identifies variable values and types in the second log message (i.e., \textit{username} and \textit{uid}). However, it cannot recognize the whole address of \logtext{video.5054399.com:80} as one variable in the first log message.

\subsection{The Performance of ChatGPT-based Log Parsing under Different Prompting Methods}

\subsubsection{With few-shot scenarios}
We evaluate the performance of ChatGPT in few-shot scenarios using PT\ref{pt:fewshot}. For the 1-shot scenario, we search for the most frequent log message and use it as the example for ChatGPT. For 2-shot and 4-shot scenarios, we apply a few-shot random sampling algorithm~\cite{le2023log} to select 2 and 4 examples for ChatGPT.
Table~\ref{tab:rq2-few-shot} shows the results.
We observe a noticeable improvement of 19.5\% and 29.9\% in MLA and ED, respectively, with just one example of log parsing.
With 4 examples, ChatGPT achieves the best MLA and ED on all 16 datasets and significantly outperforms the other log parsers. It also achieves a comparable GA to the second best log parser, SPINE.
The results indicate that ChatGPT is able to learn log parsing from a few demonstrations and achieves good performance. It also shows that ChatGPT exhibits good generality to a variety of log data through few-shot learning.

\begin{table}[h]
\caption{The results in few-shot scenarios}
\label{tab:rq2-few-shot}
\vspace{-6pt}
\centering
\footnotesize
\setlength{\tabcolsep}{2.5pt}
\renewcommand{\arraystretch}{1.2}
\resizebox{\linewidth}{!}{
\begin{tabular}{c|ccc|ccc|ccc}
\hline
\multirow{2}{*}{Dataset} & \multicolumn{3}{c|}{\textbf{1-shot}} & \multicolumn{3}{c|}{\textbf{2-shot}} & \multicolumn{3}{c}{\textbf{4-shot}} \\ \cline{2-10} 
 & GA & MLA & ED & GA & MLA & ED & GA & MLA & ED \\ \hline
HDFS & 0.903 & 0.939 & 0.063 & 0.960 & 0.993 & 0.007 & 0.845 & 0.993 & 0.105 \\
Hadoop & 0.787 & 0.588 & 11.629 & 0.959 & 0.600 & 5.448 & 0.969 & 0.623 & 4.941 \\
Spark & 0.910 & 0.880 & 1.186 & 0.873 & 0.865 & 2.080 & 0.887 & 0.925 & 0.622 \\
Zookeeper & 0.842 & 0.666 & 1.500 & 0.779 & 0.663 & 1.588 & 0.842 & 0.545 & 2.021 \\
BGL & 0.888 & 0.919 & 1.648 & 0.936 & 0.934 & 2.739 & 0.952 & 0.935 & 2.962 \\
HPC & 0.872 & 0.897 & 1.029 & 0.930 & 0.935 & 0.675 & 0.932 & 0.938 & 0.461 \\
Thunderb & 0.172 & 0.492 & 6.821 & 0.575 & 0.827 & 5.810 & 0.473 & 0.791 & 2.938 \\
Windows & 0.567 & 0.638 & 6.853 & 0.566 & 0.483 & 9.120 & 0.982 & 0.979 & 0.727 \\
Linux & 0.620 & 0.671 & 2.507 & 0.753 & 0.718 & 2.209 & 0.742 & 0.719 & 2.253 \\
Android & 0.810 & 0.663 & 9.726 & 0.870 & 0.682 & 11.932 & 0.884 & 0.698 & 7.384 \\
HealthApp & 0.908 & 0.657 & 5.952 & 0.920 & 0.742 & 3.408 & 0.920 & 0.747 & 3.211 \\
Apache & 0.731 & 0.946 & 0.486 & 0.731 & 0.793 & 1.863 & 1 & 1 & 0 \\
Proxifier & 0 & 0.329 & 6.621 & 0.024 & 0.315 & 11.631 & 0.050 & 0.781 & 2.379 \\
OpenSSH & 0.240 & 0.374 & 4.742 & 0.544 & 0.209 & 5.860 & 0.523 & 0.512 & 4.210 \\
OpenStack & 0.152 & 0.389 & 8.252 & 0.343 & 0.434 & 5.267 & 0.513 & 0.958 & 0.527 \\
Mac & 0.577 & 0.342 & 27.687 & 0.653 & 0.503 & 16.486 & 0.670 & 0.500 & 15.166 \\ \hline
Average & 0.623 & 0.649 & 6.044 & 0.713 & 0.668 & 5.383 & 0.761 & 0.790 & 3.119 \\ \bottomrule
\end{tabular}
\vspace{-6pt}
}
\end{table}


\subsubsection{With different prompts}
Different prompts could lead to different results when applying LLM. In this RQ, we evaluate the performance of ChatGPT on log parsing using (1) \textit{a simple prompt} (PT\ref{pt:zero-shot-simple}): we directly ask ChatGPT to return the template of a log message; 
and (2) \textit{an enhanced prompt} (PT\ref{pt:zero-shot-enhance}): we specifically ask ChatGPT to follow three steps of log parsing: identify variables, abstract variables, and output a static template. Table~\ref{tab:rq2-prompt} shows the results.

\begin{table}[h]
\caption{The results with different prompts}
\label{tab:rq2-prompt}
\vspace{-6pt}
\centering
\footnotesize
\setlength{\tabcolsep}{2.5pt}
\renewcommand{\arraystretch}{1.2}
\resizebox{\linewidth}{!}{
\begin{tabular}{c|ccc|ccc|ccc}
\toprule
\multirow{2}{*}{\textbf{Dataset}} & \multicolumn{3}{c|}{\textbf{PT1}} & \multicolumn{3}{c|}{\textbf{PT3 (Simple)}} & \multicolumn{3}{c}{\textbf{PT4 (Enhance)}} \\ \cline{2-10} 
 & GA & MLA & ED & GA & MLA & ED & GA & MLA & ED \\ \hline
HDFS & 0.960 & 0.939 & 0.062 & 0.413 & 0.884 & 0.535 & 0.920 & 0.892 & 1.197 \\
Hadoop & 0.795 & 0.525 & 11.017 & 0.740 & 0.450 & 11.556 & 0.801 & 0.449 & 10.709 \\
Spark & 0.925 & 0.922 & 0.596 & 0.623 & 0.788 & 0.880 & 0.700 & 0.922 & 0.662 \\
Zookeeper & 0.667 & 0.233 & 5.460 & 0.797 & 0.233 & 6.672 & 0.648 & 0.273 & 4.409 \\
BGL & 0.878 & 0.790 & 5.258 & 0.243 & 0.686 & 8.512 & 0.947 & 0.863 & 3.329 \\
HPC & 0.807 & 0.497 & 3.498 & 0.592 & 0.605 & 5.277 & 0.920 & 0.908 & 0.816 \\
Thunderb & 0.568 & 0.808 & 5.933 & --- & --- & --- & 0.255 & 0.505 & 3.395 \\
Windows & 0.686 & 0.301 & 17.623 & 0.148 & 0.292 & 20.239 & 0.403 & 0.525 & 8.602 \\
Linux & 0.910 & 0.635 & 3.328 & 0.286 & 0.657 & 3.428 & 0.445 & 0.594 & 3.448 \\
Android & 0.711 & 0.549 & 10.763 & 0.754 & 0.574 & 12.087 & 0.922 & 0.652 & 7.349 \\
HealthApp & 0.898 & 0.628 & 6.560 & 0.767 & 0.637 & 6.498 & 0.886 & 0.636 & 6.425 \\
Apache & 1 & 1 & 0 & 0.984 & 0.708 & 4.955 & 1 & 1 & 0 \\
Proxifier & 0.001 & 0.014 & 27.025 & 0 & 0.001 & 18.424 & 0.001 & 0.016 & 27.730 \\
OpenSSH & 0.659 & 0.170 & 7.854 & 0.261 & 0.335 & 6.609 & 0.462 & 0.451 & 4.837 \\
OpenStack & 0.449 & 0.433 & 7.440 & 0.355 & 0.315 & 10.670 & 0.524 & 0.433 & 7.004 \\
Mac & 0.619 & 0.248 & 25.530 & 0.434 & 0.228 & 39.599 & 0.614 & 0.380 & 17.919 \\ \hline
Average & 0.721 & 0.543 & 8.621 & 0.493 & 0.493 & 10.396 & 0.653 & 0.594 & 6.739 \\ \bottomrule
\end{tabular}
}
\vspace{-6pt}
\end{table}

We notice that with a simple prompt, ChatGPT can hardly understand the concept of log parsing and thus achieve low accuracy (e.g., 0.493 GA). There are many cases where ChatGPT asks for more information when we use PT\ref{pt:zero-shot-simple}. In contrast, with the aid of the enhanced prompt (PT\ref{pt:zero-shot-enhance}), ChatGPT can perform log parsing more effectively. Specifically, it achieves an improvement of 9.4\% and 21.8\% over PT\ref{pt:zero-shot-1} in MLA and ED, respectively. Overall, the design of prompts has a large impact on the performance of log parsing with ChatGPT. Including a clearer intention in the prompt could enhance parsing accuracy.
\section{Discussion}



Based on our findings, we highlight several challenges and prospects for ChatGPT-based log parsing.



\textbf{(1) Handling log-specific data}.
Our study on log parsing indicates that it is promising to analyze log data with ChatGPT. However, it also shows that ChatGPT still faces difficulties in recognizing log-specific information generated during runtime (e.g., domain URLs, API endpoint addresses, etc.). Since these information occur frequently in log data, it could hinder the ability of ChatGPT in understanding log data.

\textbf{(2) The selection of demonstrations}. Our experimental results show that ChatGPT exhibits good performance with few-shot prompting. Overall, the performance of ChatGPT can be improved with more demonstrations. However, we observe that in some cases, these demonstrations could bring noise and confuse the model (see Table~\ref{tab:rq2-few-shot}). Therefore, it is necessary to ensure the quality of selected demonstrations. How to select a small yet effective set of examples is an important future work. 

\textbf{(3) Designing better prompts.} In this paper, we found that different prompts could have a big impact on the performance of ChatGPT-based log parsing. Although many prompting methods have been proposed~\cite{liu2023pre, weichain, wang2023plan},
it remains to explore which prompting method is suitable for log parsing, how to systematically design prompts, and whether there are better prompting methods. Future work is required toward designing better prompts for log parsing.

\textbf{(4) Toward semantic-aware log parsing.} We observe that ChatGPT is able to not only extract the template associated with variables but also semantically identify the categories of variables (see Figure~\ref{fig:chatgpt-examples}).
This awareness of variables' semantics could improve the accuracy of downstream tasks such as anomaly detection~\cite{li2023did, huo2023semparser}. Although achieving good initial results, future studies should be conducted to comprehensively evaluate the ability of ChatGPT toward semantic-aware log parsing.





\section{Conclusion}
The paper discusses the potential of using ChatGPT, a popular large language model, for log parsing. We have designed appropriate prompts to guide ChatGPT to understand the log parsing task and compared its performance with state-of-the-art log parsers in zero-shot and few-shot scenarios. Our experimental results show that ChatGPT can achieve promising results for log parsing with appropriate prompts, especially with few-shot prompting. We also outline several challenges and opportunities for ChatGPT-based log parsing. In our future work, we will comprehensively evaluate the performance of ChatGPT and other LLMs on more log analytics tasks.


Our experimental data are available at: \url{https://github.com/LogIntelligence/log-analytics-chatgpt}.

\section*{Acknowledgment}
This work is supported by Australian Research Council (ARC) Discovery Projects (DP200102940, DP220103044).
We also thank anonymous reviewers for their insightful and constructive comments, which significantly improve this paper.

\bibliographystyle{IEEEtran}
\bibliography{ref}
\balance

\end{document}